\documentclass[fleqn,usenatbib]{mnras}

\usepackage{newtxtext,newtxmath}
\usepackage[T1]{fontenc}

\DeclareRobustCommand{\VAN}[3]{#2}
\let\VANthebibliography\thebibliography
\def\thebibliography{\DeclareRobustCommand{\VAN}[3]{##3}\VANthebibliography}

\usepackage{graphicx}	% Including figure files
\usepackage{amsmath}	% Advanced maths commands
\usepackage{comment}
\usepackage{multirow}
\usepackage{hyperref}
\usepackage{chemformula}
\usepackage{xcolor}
\newcommand{\orcid}[1]{\href{https://orcid.org/#1}{\hskip2pt\includegraphics[width=9pt]{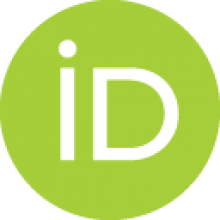}}}
\title[Adsorption of S-bearing species on olivine]{ Theoretical Modeling of the Adsorption of Neutral and Charged Sulphur-Bearing Species onto Olivine Nanoclusters}

\author[J. Perrero et al.]{
Jessica Perrero\orcid{0000-0003-2161-9120},$^{1,2}$
Leire Beitia-Antero\orcid{0000-0003-0833-4075},$^{4,5}$
Asunción Fuente\orcid{0000-0001-6317-6343},$^{3}$
Piero Ugliengo\orcid{0000-0001-8886-9832}$^{2}$
\thanks{E-mail: piero.ugliengo@unito.it}
and Albert Rimola\orcid{0000-0002-9637-4554},$^{1}$
\thanks{E-mail: albert.rimola@uab.cat}
\\
\\
$^{1}$Departament de Qu\'{i}mica, Universitat Aut\`{o}noma de Barcelona, Bellaterra, 08193, Catalonia, Spain\\
$^{2}$Dipartimento di Chimica and Nanostructured Interfaces and Surfaces (NIS) Centre, Universit\`{a} degli Studi di Torino, via P. Giuria 7, 10125, Torino, Italy\\
$^{3}$Centro de Astrobiolog\'{\i}a (CSIC/INTA), Ctra.deTorrej\'on a Ajalvir km 4, 28806, Torrej\'on de Ardoz, Spain\\
$^{4}$AEGORA Research Group - Joint Center for Ultraviolet Astronomy, Universidad Complutense de Madrid, Plaza de Ciencias 3, 28040 Madrid, Spain \\
$^{5}$Departamento de Estad\'{\i}stica e Investigaci\'on Operativa, Fac. de CC. Matem\'aticas, Plaza de Ciencias 3, 28040 Madrid, Spain
}
\date{Accepted XXX. Received YYY; in original form ZZZ}

\pubyear{2023}

\begin{document}
\label{firstpage}
\pagerange{\pageref{firstpage}--\pageref{lastpage}}
\maketitle

\begin{abstract}

%MESSAGE OF THE PAPER: the BE are so high that if the species fall on the grain they will be buried there and not desorb back into the gas phase.

Sulphur depletion in the interstellar medium (ISM) is a long-standing issue, as only 1\% of its cosmic abundance is detected in dense molecular clouds (MCs), while it does not appear to be depleted in other environments. In addition to gas phase species, MCs also contain interstellar dust grains, which are irregular, micron-sized, solid aggregates of carbonaceous materials and/or silicates. Grains provide a surface where species can meet, accrete, and react.
Although freeze-out of sulphur onto dust grains could explain its depletion, only OCS and, tentatively, SO$_2$ were observed on their surfaces. Therefore, it is our aim to investigate the interaction between sulphur-containing species and the exposed mineral core of the grains at a stage prior to when sulphur depletion is observed.
Here, the grain core is represented by olivine nanoclusters, one of the most abundant minerals in the ISM, with composition Mg$_4$Si$_2$O$_8$ and Mg$_3$FeSi$_2$O$_8$. We performed a series of quantum mechanical calculations to characterize the adsorption of 9 S-bearing species, both neutral and charged, onto the nanoclusters. Our calculations reveal that the \ch{Fe\bond{single}S} interaction is preferred to \ch{Mg\bond{single}S}, causing sometimes the chemisorption of the adsorbate. These species are more strongly adsorbed on the bare dust grain silicate cores than on water ice mantles, and hence therefore likely sticking on the surface of grains forming part of the grain core. This demonstrates that the interaction of bare grains with sulphur species in cloud envelopes can determine the S-depletion observed in dense molecular clouds.

\end{abstract}


%%%%%%%%%%%%%%%%%%%%%%%%%%%%%%%%%%%%%%%%%%%%%%%%%%

%%%%%%%%%%%%%%%%% BODY OF PAPER %%%%%%%%%%%%%%%%%%

\section{Introduction}

On Earth, sulphur is one of the key elements that contributes to the engine of life. It is an essential component of our bodies, as it is found in amino acids such as cysteine and methionine, in coenzyme A, in vitamin B1 and biotin, in antioxidants as glutathione and in many more molecules \citep{voet2008}. However, sulphur played an important role long before life appeared on Earth, and its chemistry in the interstellar medium (ISM), the environment where stars and planets are formed, is nowadays a challenge for the astronomical and the astrochemical community. 
Sulphur is the tenth most abundant element in the Universe. Its cosmic abundance, [S]/[H] = 1.8*10$^{-5}$ \citep{anders1989}, equal to an S/O ratio of 1/37, is used as a reference for measuring sulphur abundance during the several stages of the evolution of a planetary system. While in diffuse clouds the majority of sulphur resides in the gas phase as S$^+$ \citep{jenkins2009}, it is not clear which forms it assumes in translucent and dense clouds (also called molecular clouds, MCs) \citep{woods2015,laas2019}. \cite{hilyblant2022} suggests that in chemically young cores, atomic sulphur is still the main carrier, which however could not be directly observed. Moreover, it would gradually deplete as the cloud evolves, as also inferred in \cite{fuente2023}. The phenomenon of freeze-out that occurs in MCs could be responsible for the adsorption of gas-phase sulphur onto the surface of icy grains \citep{caselli1994}, a case in which the presence of H$_2$S on their surface would be expected. However, only OCS and tentatively SO$_2$ were observed in MCs and young stellar object (YSOs) ices \citep{boogert2015,boogert2022,mcclure2023} and only upper limits have been estimated for H$_2$S \citep{smith1991,mcclure2023}. More recent studies suggest that sulphur is mostly contained in organic species \citep{laas2019} or in a refractory residue composed of species characterized by the \ch{S\bond{single}S} bond (like H$_2$S$_n$ with 3 $\le$ n $\le$ 8, and chains of S$_n$ with 2 $\le$ n $\le$ 3), along with complex species such as hexathiepan (S$_6$CH$_2$), which are products of ice processing, as experiments indicate \citep{ferrante2008,jimenez2011,cazaux2022}. Nevertheless, numerous sulphur-bearing species (S$_2$, S$_3$ and S$_4$, CH$_3$SH, C$_2$H$_6$SH, together with the most common H$_2$S, OCS, SO, S$_2$, SO$_2$ and CS$_2$) are detected in comets as 67P/Churyumov-Gerasimenko \citep{calmonte2016}, whose composition seems to be inherited from the prestellar and protostellar evolutionary phases \citep{drozdovskaya2019}. Comets contain a record of the chemical composition of the primitive Solar Nebula at the place where they formed, 4.6 Gyr ago \citep{altwegg2019}. The same applies to a particular type of meteorites, in which sulphur is found in the form of minerals containing iron and nickel \citep{jm2012meteoritos}. Among the variety of existing meteorites, chondrites represent fairly well the material that surrounded the young Sun and from which the planets formed. 
%Chondrites are classified in ordinary, enstatite, carbonaceous and anomalous ones. 
Ordinary chondrites contain S mostly in the form of troilite FeS (a variety of pyrrhotite, Fe$_{(1-x)}$S) \citep{kallemeyn1989}, while carbonaceous chondrites contain also pentlandite (Fe,Ni)$_9$S$_8$, and some inorganic sulphates and aromatic organic compounds. \citep{gao1993,jm2012meteoritos,yabuta2007} On the other hand, enstatite chondrites are rich in sulphides like niningerite, alabandite and other alkaline sulphides \citep{sears1982}.

Thus, there seems to be a missing piece of the puzzle between the gas-phase S-bearing species present in diffuse clouds and the solid sulphides found in the remnants of the protoplanetary disks. %meteroites and comets 
To understand the fate of sulphur, we aim to investigate what can happen in diffuse clouds, when some simple sulphur species interact with bare dust grains, a system that already contains the atomic ingredients necessary to form the sulphides observed in more evolved environments. Depending on the C/O ratio of the AGB stars where they are formed, dust grains are formed by carbonaceous materials or silicates and represent 1\% of the ISM mass \citep{potapov2021}. The main families of interstellar silicates are pyroxenes and olivines with chemical composition
Mg$_x$Fe$_{(1-x)}$SiO$_3$ and Mg$_{2x}$Fe$_{(2-2x)}$SiO$_4$ (with x = 0–1) \citep{henning2010}. These nanometre to micrometre-sized nonspherical, porous dust particles lock up nearly 30\% of oxygen and 100\% of silicon, magnesium and iron available in the interstellar medium \citep{vandishoeck2013}. 
Commonly, interstellar grain silicates are structurally amorphous, although Mg-rich crystalline silicates have also been observed around young stars \citep{molster2005,spoon2022}. 

In this work, we modelled dust grains as olivine nanoclusters, as in a recent work by \cite{peralta2022}. In the latter, both cluster and periodic approaches were considered to study the formation of water, and were previously adopted by \cite{navarro2014olivine,navarro2016} to study the formation of H$_2$. To our knowledge, the only study of S-bearing species interacting with an olivine surface is a paper by some of us \citep{rimola2017} in which butan-1-sulphonic acid was adsorbed on the (010) surface of forsterite (Mg$_2$SiO$_4$) to seek the correlation between binding energies and measured abundances of a class of soluble organic compounds found in carbonaceous meteorites. Here, we decided to focus on the cluster approach, firstly developing a reliable computational methodology to simulate olivine clusters, using here the adsorption of H$_2$S as a test case. We then characterized the adsorption on the Mg and Fe sites of olivine clusters of other sulphur bearing species, both neutral and charged: S, S$^+$, CS, SH, SH$^+$, SO, SO$^+$, H$_2$S, and H$_2$S$^+$. For each species, we computed the reaction energy, which can be recast as interaction energy, $\Delta$E, which is a negative quantity for an exothermic reaction. Usually, in the context of astrochemical numerical models, it is the binding energy BE (BE=-$\Delta$E) the quantity that defines the strength of the interaction between the species and the surface. The BE, as defined above, is computed by using the electronic energies of the involved species, assuming that the nuclei are immobile. Quantum mechanics imposes that even at 0 K the nuclei undergo a vibrational motion, which contributes to the electronic energy through the zero point energy (ZPE). It is, therefore the BE(0) = BE $-$ $\Delta$ZPE, the key parameter to be used in astrochemical models, determining whether desorption phenomena can take place \citep{penteado2017}, and also governing the diffusion process, as diffusion barriers are usually assumed to be a fraction of the BE(0) \cite[e.g.,][]{karssemeijer2014,cuppen2017,kouchi2020,mate2020}.

The work is organized as follows: in Section \ref{sec:methodology} we describe the methodology used in the study, Section \ref{sec:risultati} contains the results, and in Section \ref{sec:discussione} we discuss our finding and their implications in the sulphur depletion problem. Finally, Section \ref{sec:conclusioni} summarizes the conclusions.

\section{Methods}\label{sec:methodology}

\begin{figure}
    \centering
    \includegraphics[width=\columnwidth]{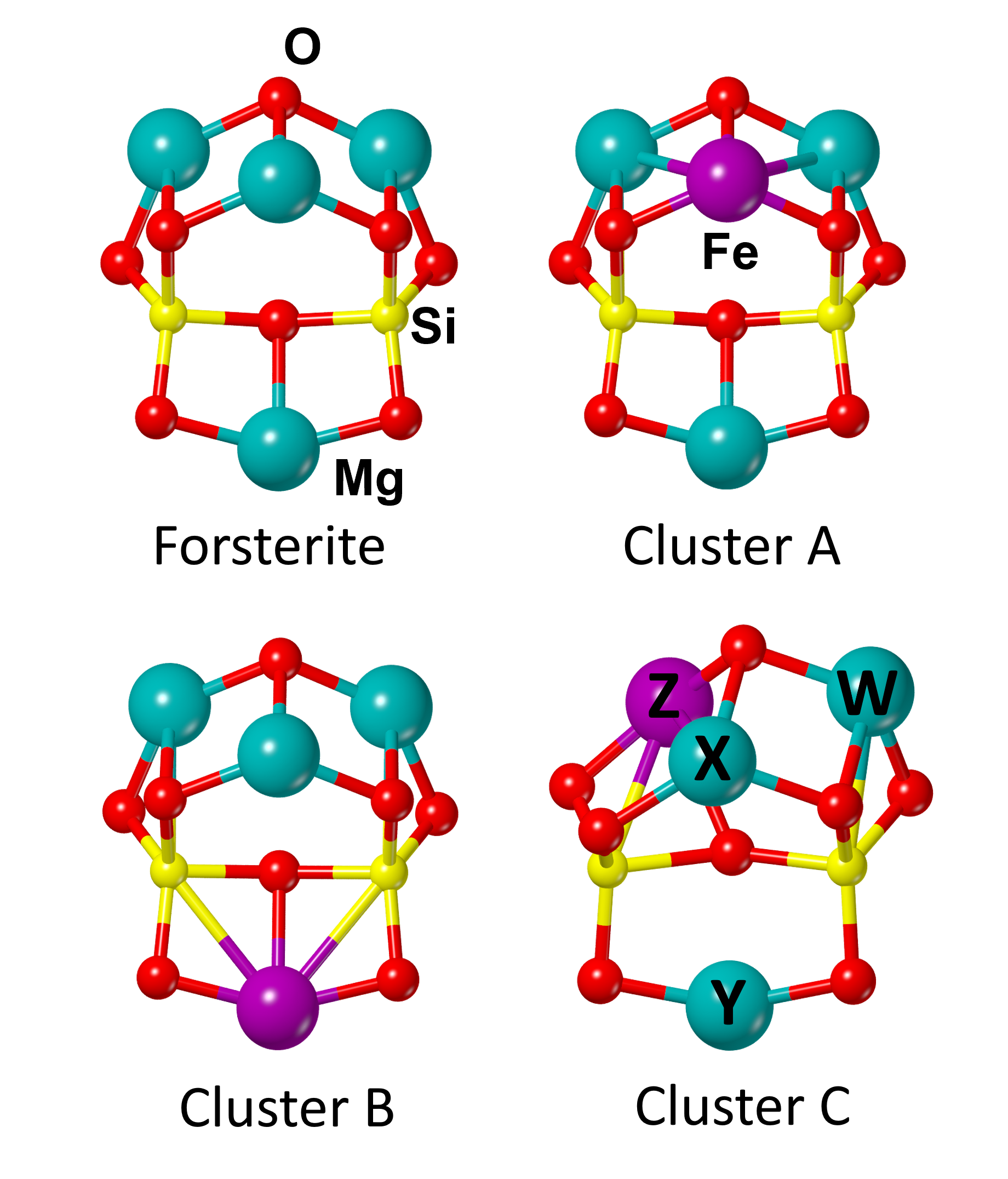}
    \caption{Forsterite and olivine clusters used in this work as dust grains models. The labels in cluster C are useful to identify adsorption sites.}
    \label{fig:clusters}
\end{figure}
 
The olivine clusters presented in \cite{peralta2022} and their forsterite parent cluster developed by \cite{Macia-EScatllar2019} were adopted to model the core of a dust grain. Each model consists of a cluster of two silicate units with composition Mg$_{(4-x)}$Fe$_{x}$Si$_2$O$_8$ (with x = 0–1). Clusters A, B, and C are characterized by the presence of a Fe atom at positions X, Y, and Z, respectively (see Figure \ref{fig:clusters}). Please, note that W and Z positions are equivalent by symmetry. The free Fe$^{2+}$ electronic valence configuration is 4s$^0$ 3d$^6$, allowing the system to exist in three different spin states: singlet, triplet, and quintet. As reported in the literature \citep{navarro2014olivine,navarro2016}, the most stable configuration is that showing the maximum spin multiplicity, therefore the quintet, followed by the triplet and the singlet states at higher energies. 
%The same holds true for our calculations.

We first performed a benchmark study with the aim of finding a suitable methodology to describe the adsorption of S-bearing species onto olivine clusters. All the calculations were run with the ORCA program v.5.0.3 \citep{neese2022}. We optimized (using default settings) clusters A, B, and C in the quintet, triplet and singlet spin states with: i) the methodology adopted in \cite{peralta2022}, namely B3LYP-D3(BJ) \citep{lee1988,becke1988,becke1993} combined with the 6-311++G(d,p) basis set \citep{hehre1972,hariharan1973}; ii) the meta-generalized-gradient approximation r$^2$SCAN-3c, which goes with its own tailored triple zeta quality basis set \citep{grimme2021}, iii) the range-separated hybrid meta-GGA density functional $\omega$B97M-V \citep{mardirossian2017}, and iv) BHLYP-D3, a hybrid density functional which features 50\% of exact Hartree-Fock exchange \citep{bhlyp-becke1993,lee1988}. The last two functionals were combined with the def2-TZVP basis set \citep{weigend2005}. The calculations have been performed with the spin-unrestricted formalism in case of open-shell systems \citep{pople:1995}. We compared the values obtained with these DFT functionals against the DLPNO-CCSD(T)/aug-cc-pVTZ \citep{guo2018} single energy calculations performed on each optimized geometry. The calculations were carried out with a tight-PNO setup and the default settings for the SCF.

The second part of the benchmark involved the characterization of the adsorption of H$_2$S and H$_2$O on the forsterite grain nanoclusters, starting from four complexes manually built to allow the two species to interact with sites X, W, Y and with neighbouring oxygen atoms of site W (which as mentioned above is equal to site Z). Adsorbing H$_2$S with r$^2$SCAN-3c and B3LYP-D3(BJ) results in some cases in its spontaneous dissociation. Thus, we used the Nudged Elastic Band (NEB) algorithm to find the minimum energy path connecting the reactant (physisorbed complex) and the product (chemisorbed complex). The algorithm generates a number of configurations of the atoms, referred to as images of the system (in our case we chose 8 images) that link the two minima with the aim of finding the image with the highest energy. Subsequently, the latter structure is optimized as a transition state following the eigenvector associated with the eigenvalue of the Hessian matrix.

Once we selected an appropriate level of theory, we initially characterized the binding sites of the four olivine clusters by adsorbing H$_2$S, H$_2$O, CS and CO. Because of symmetry, we identified three adsorption sites (X, W, Y) for forsterite cluster, cluster A and cluster B, while in cluster C the Fe substitution at position Z causes a loss of symmetry and the emergence of an additional binding site. We then characterized the adsorption of the nine S-bearing species on cluster C, which is the most stable structure and presents the largest number of binding sites. Each species was described in its ground state: CS, H$_2$S, and SH$^+$ as singlets, S$^+$, SH, SO$^+$, and H$_2$S$^+$ as doublets, and finally S and SO as triplets. Combining the spin states of the S-bearing species with the ground state of cluster C, i.e., the quintet, we obtained adsorption complexes of CS, H$_2$S, and SH$^+$ in the quintet state, S$^+$, SH, SO$^+$, and H$_2$S$^+$ in the sextet state and S and SO in the septet state. For each complex, the full set of harmonic frequencies was computed to establish each structure to be a true minimum on the potential energy surface (PES). From them, the zero point vibrational energy (ZPE) correction was computed to correct the binding energy, BE, a quantity defined as the opposite of the interaction energy, obtained with the equation:
\begin{equation}
    BE = - \Delta E_{int} = - E_{complex} + E_{cluster} + E_{adsorbate}
\end{equation}
to be corrected for the ZPE as:
\begin{equation}
BE(0) = BE - \Delta ZPE 
\end{equation}
where $\Delta$ZPE = ZPE$_{complex}$ - ZPE$_{cluster}$ - ZPE$_{adsorbate}$.

%%%%%%%%%%%%%%%%%%
% We did not compute the BSSE correction for each structure, given that the def2-tzvp basis set is large enough to avoid a large error.
%%%%%%%%%%%%%%%%%%%

\section{Results}\label{sec:risultati}

\subsection{Benchmark}

Clusters A, B, and C were optimized to characterize their stability from both a structural and an electronic point of view.  
Data on the quintet and the singlet states are reported in Table \ref{tab:benchmark}, while those concerning the triplet state were excluded due to some convergence problems that arose during the optimization of the structures. We did not aim to compare the energy difference between clusters of different spin states, as we know that DFT is not reliable for treating systems in their excited states, but we compared the relative energies between different clusters with the same spin multiplicity ($\Delta$E). Cluster C was the most stable of the group and it was taken as a reference to calculate the $\Delta$E of cluster A and B. Among the four functionals tested in the benchmark,  $\omega$B97M-V showed the smallest error when compared to DLPNO-CCSD(T), with an average error of 11.1\% for the quintet and of 21.9\% for the singlet spin states. Although the percentage error appeared to be large, the absolute errors corresponded to a few kJ mol$^{-1}$, which are always < 4 kJ mol$^{-1}$ for $\omega$B97M-V, therefore, in the limit of chemical accuracy. BHLYP-D3(BJ) performed similarly to $\omega$B97M-V, followed by B3LYP-D3(BJ), while r$^2$SCAN-3c, lacking of Hartree-Fock exchange, was the less suitable functional. However, it is worth pointing out the surprising performance of r$^2$SCAN-3c in the description of the quintet spin state, despite the presence of unpaired electrons, with an error of only 9.3\%. Nevertheless, the representation of the singlet spin state was not satisfying, showing an error of 72.9\%.
%(1 kcal/mol = 4.184 kJ/mol)
In general, the energies of the clusters in the singlet spin state were affected by larger errors than those of the structures in the quintet ground state, confirming that DFT does not represent a suitable choice when describing excited states. Nevertheless, the stability order documented in the literature was well reproduced with every functional adopted in the benchmark. In terms of the description of the band gap of the system, the outcome largely depends on the percentage of Hartree-Fock (HF) exchange included in the definition of the functional. Indeed, for r$^2$SCAN-3c, in which HF exchange is not accounted for, the band gap was almost zero, sometimes responsible for convergence problems, as in the case of the triplet spin state.\\

%The band gap computed with HF gives 10.9 eV. 

To expand our benchmark, we considered the adsorption of H$_2$S and H$_2$O on the forsterite (Fo) cluster. The two species were manually placed on the Fo and the obtained complexes were optimized with $\omega$B97M-V, B3LYP-D3 and r$^2$SCAN-3c functionals. When comparing the interaction energies computed with the three different functionals against their respective DLPNO-CCSD(T) single energy points, $\omega$B97M-V/def2-TZVP resulted again in the best performing level of theory (see Table \ref{tab:bench_ads}). In this case, the percentage errors were much lower than in the previous benchmark due to the larger adsorption energies involved compared to $\Delta$E. However, at least for $\omega$B97M-V, the absolute error remained below 4 kJ mol$^{-1}$. We also point out the great improvement in the performance of r$^2$SCAN-3c, justified by the absence of unpaired electrons in the system.

\begin{figure*}
    \centering
    \includegraphics[width=\linewidth]{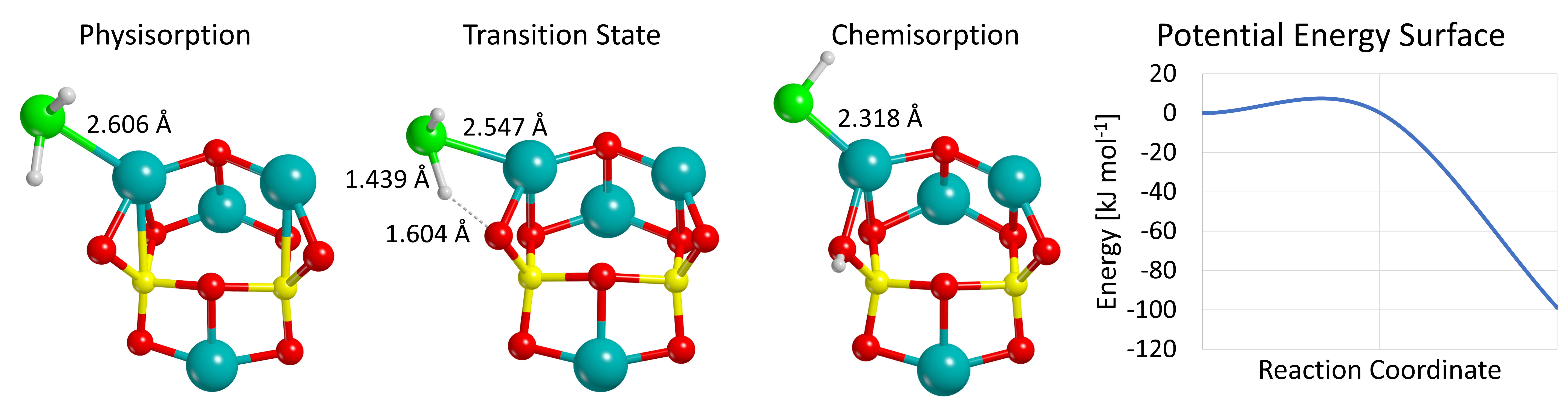}
    \caption{The dissociation of H$_2$S on the forsterite cluster (complex III) characterized at the $\omega$B97M-V/def2-TZVP level of theory. It exhibits a small potential energy barrier and becomes barrierless when taking into account the ZPE corrections.}
    \label{fig:dissociation}
\end{figure*}

When modelling the adsorption of H$_2$S on Fo, we noticed an interesting phenomenon. A spontaneous dissociation of H$_2$S was observed in complexes III and IV at r$^2$SCAN-3c level of theory, while only complex IV caused the dissociation of the adsorbate at B3LYP-D3(BJ) level of theory, and no dissociation occurred with the $\omega$B97M-V functional. In these two complexes (III and IV), H$_2$S interacts with the Mg cation located in sites W and X, respectively. The dissociated complexes obtained with r$^2$SCAN-3c are stable enough, so that a subsequent optimization with $\omega$B97M-V did not change their structure. The chemisorbed structures were more stable than the physisorbed ones by about 100 kJ mol$^{-1}$. 
Thus, we computed the energy barrier for the dissociation process at $\omega$B97M-V (that in which spontaneous dissociation is not observed), obtaining a value of 0.2 kJ mol$^{-1}$ for complex III and 7.2 kJ mol$^{-1}$ for complex IV. When considering the ZPE correction, the dissociation barriers at 0 K became -3.4 kJ mol$^{-1}$ and 3.9 kJ mol$^{-1}$ for complexes III and IV, respectively. The frequency associated with the imaginary mode of the transition state is 231\textit{i} cm$^{-1}$ for complex III and 192\textit{i} cm$^{-1}$ for complex IV. In Figure \ref{fig:dissociation}, the reactant, the transition state structure and the product are shown in the case of complex III, together with a plot of the PES along the reaction coordinate. The dissociation barriers are indeed very small, in one case even submerged below zero, indicating that the cleavage of H$_2$S into HS$^-$ and H$^+$ is likely to occur onto Fo cluster, once the molecule is adsorbed on the Mg sites. When adsorbing \ch{H2O} on Mg, we observed the spontaneous dissociation of the \ch{O\bond{single}H} bond in complex IV, in which the oxygen atom interacts with the Mg cation of site X at both r$^2$SCAN-3c and B3LYP-D3(BJ) levels of theory. We did not computed the PES of the dissociation at $\omega$B97M-V level of theory, however, we assume that a small barrier for the process is present, as for \ch{H2S}. Thus, we can conclude that the dissociation barrier depends on the methodology and that can either be submerged below zero or be a few kJ mol$^{-1}$. The two benchmark studies indicate $\omega$B97M-V/def2-TZVP to be the best performing methodology with an average error of 16.5\% on the relative stability of olivine clusters and of 2.6\% on H$_2$S and H$_2$O adsorption energies, becoming this functional our choice for the following calculations. 

\subsection{Cluster versus periodic approach}
The computational approach adopted in this work to study the interaction of S-bearing species with the silicate core of the grains may appear too simplistic due to the limited size of the cluster models. The exposure of the metal centres could be responsible for their higher reactivity, which could influence the outcomes of the calculations. 
Enlarging the cluster model to increase the coordination of the metal centres is not practical due to the steep increase of the computational cost of the calculations. Therefore, we adopted a periodic approach to model different extended forsterite slab surfaces, with the closest level of theory to that of the cluster. The periodic approach represents the opposite situation compared to the cluster, as the metal center binding sites (albeit at the surfaces) possess the highest coordination due to the extended nature of the slabs.
The cluster-based adsorption complexes described in Table \ref{tab:bench_ads} were optimized at PBE level of theory, combined with the def2-TZVP basis set. For the periodic calculations, \ch{H2S} was adsorbed onto the (001), (021), (101), (110), (111), and (120) forsterite surfaces, which were previously characterized in \cite{zamirri2017} and \cite{bancone2023}. The periodic adsorption complexes were also optimized at PBE, and the forsterite atoms were described with the basis set proposed by \cite{bruno2014}, while \ch{H2S} with the same def2-TZVP basis set adopted for the cluster calculations.  
We did not include dispersion interactions in the description of the system, as it was not possible to reproduce the desired level of theory (D2 with customized dispersion coefficient for the Mg atom) in both the cluster and the periodic calculations due to specific implementation of the D2 coding in the ORCA program.
Despite the missing dispersion term, the calculations show that \ch{H2S} still spontaneously dissociates on the Fo cluster, in complexes I, II, III and IV. On the periodic systems, molecular physisorption is found on the most stable forsterite surfaces, (120) and (001), while on the less stable (101), (111), (021), and (110), \ch{H2S} dissociative chemisorption takes place.
Therefore, we can thus conclude that H$_2$S chemisorption is not simply an artifact of the size of the cluster that exposes low-coordinated metal centres. Depending on the surface and on the specific adsorption site, it is possible to have spontaneous dissociation of \ch{H2S} also on large and extended surfaces in which the ion coordination is more complete than in the cluster. We remark that the level of theory chosen to characterize the system may be responsible for the presence or absence of a small barrier for the dissociation process.

\begin{table*}
   \caption {Relative energies ($\Delta$E in the text) of olivine clusters optimized with different DFT functionals and their DLPNO-CCSD(T) single energy point calculations. Cluster C is the most stable one and is taken as a reference to compute the $\Delta$E of clusters A and B.}
    \centering
\begin{tabular}{lcccccccc} 
\hline
\multirow{2}*{Quintet}  & \multirow{2}*{r$^2$SCAN-3c}	 & DLPNO// &  \multirow{2}*{$\omega$B97M-V} & DLPNO// & \multirow{2}*{BHLYP-D3(BJ)} & DLPNO// & \multirow{2}*{B3LYP-D3(BJ)} & DLPNO// \\
& & r$^2$SCAN-3c & & $\omega$B97M-V & & BHLYP-D3(BJ) & & B3LYP-D3(BJ) \\

\hline
Cluster A	 & 31.5 & 28.3  & 32.6 & 29.1 & 34.4 & 27.1 & 34.0 & 32.6 \\
Cluster B	 & 10.0 & 10.8 & 12.9 & 11.7 & 9.6 & 9.4 & 14.0 & 11.9 \\
Error  & 9.3\% & & 11.1\% & & 14.1\% & & 17.5\% & \\
\hline
\multirow{2}*{Singlet}  & \multirow{2}*{r$^2$SCAN-3c}	 & DLPNO// &  \multirow{2}*{$\omega$B97M-V} & DLPNO// & \multirow{2}*{BHLYP-D3(BJ)} & DLPNO// & \multirow{2}*{B3LYP-D3(BJ)} & DLPNO// \\  
& & r$^2$SCAN-3c & & $\omega$B97M-V & & BHLYP-D3(BJ) & & B3LYP-D3(BJ) \\
\hline
Cluster A	 & 18.1 & 10.5 & 13.3 & 11.5 & 7.1 & 10.8 & 14.2 & 12.1 \\
Cluster B	 & 22.9 & 13.2 & 17.5 & 13.6 & 18.3 & 13.5 & 19.8 & 13.6 \\
Error &  72.9\% & & 21.9\% & & 35.2\% & & 31.5\% & \\
\hline
Average Error  & 41.1\% & & 16.5\% & & 24.7\% & & 24.5\% & \\
\hline
\label{tab:benchmark}
\end{tabular}
\end{table*}

\begin{table*}
   \caption {Adsorption energies ($\Delta$E$_{int}$ in the text) of H$_2$S and H$_2$O on the forsterite cluster optimized with different DFT functionals and their DLPNO-CCSD(T) single energy point calculations. Complex I and II differ for the orientation of the adsorbate with respect to the binding site (\#). The asterisk (*) indicates the cases in which spontaneous dissociation takes place. The percentage error is obtained as the average over those computed for each DFT method against the DLPNO//DFT energy value. The average is done over the four complexes of both H$_2$O and H$_2$S together.}

    \centering
\begin{tabular}{lccccccc} 
\hline
	 \multirow{2}*{Adsorption of H$_2$S} & Metal  & \multirow{2}*{r$^2$SCAN-3c}	 & DLPNO// &  \multirow{2}*{$\omega$B97M-V} & DLPNO// & \multirow{2}*{B3LYP-D3(BJ)} & DLPNO// \\
& center & & r$^2$SCAN-3c & & $\omega$B97M-V & & B3LYP-D3(BJ) \\
\hline
Complex I & 13  & -79.5	& -74.5  & -74.1 &	-76.2 & -84.4	& -84.5  \\    % geometry 1
Complex II & 13$^{\#}$ & -90.9	& -92.1    & -86.6 &	-92.3 & -101.5	& -101.2  \\  % geometry 2
Complex III & 10 & -185.8*	& -176.0* & -77.0 &	-80.0 & -88.2	& -89.3  \\   % geometry 3
Complex IV & 6  & -183.0*	& -174.5* & -70.3 &	-72.8 & -194.7*	& -186.1*    \\    % geometry 4
%Error & 4.6\% & & 4.0\% & & 1.6\% & \\
\hline
	 \multirow{2}*{Adsorption of H$_2$O} & Metal  & \multirow{2}*{r$^2$SCAN-3c}	 & DLPNO// &  \multirow{2}*{$\omega$B97M-V} & DLPNO// & \multirow{2}*{B3LYP-D3(BJ)} & DLPNO// \\
& center & & r$^2$SCAN-3c & & $\omega$B97M-V & & B3LYP-D3(BJ) \\
\hline
Complex I & 13  & -119.5	 & -124.5 & -122.9	& -125.4 & -127.7	& -135.4	  \\    % geometry 1
Complex II & 13$^{\#}$  & -119.4	 & -124.6 & -124.3	& -126.5 & -126.5	& -139.0      \\   %geometry 5
Complex III & 10  & -168.1	 & -157.1 & -167.0	& -167.8 & -178.1	& -177.5	  \\ % geometry 3
Complex IV & 6 & -153.5*	 & -161.3* & -114.5	& -115.3 & -174.5*	& -172.4*	  \\   % geometry 4
%Error & 5.0\% & & 1.2\% & & 4.0\% & \\
\hline
Average Error & & 4.8\% & & 2.6\% & & 3.4\% & \\
\hline
\label{tab:bench_ads}
\end{tabular}
\end{table*}

\subsection{Characterization of the nanocluster binding sites}

We characterized the binding sites of each nanocluster (Fo and the olivines) by adsorbing H$_2$S, CS, and their oxygen analogues. H$_2$S and H$_2$O are species with an electronegative atom (S and O) that preferentially interacts with the positively charged metal center, and two hydrogen atoms that can interact via H-bond with the oxygen atoms of the silicate unit, when geometrically feasible. On the other hand, CS and CO are species whose dipole shows a concentration of electronic charge on the carbon atom, and accordingly this atom is the one interacting with the metal center.
% while oxygen and sulphur likely interact through dispersive forces with the mineral surface.
The bond length of CS in the gas phase is 1.523 \AA, and when it is adsorbed on Mg it shortens to 1.500 \AA, due to the electron donation coming from the $\sigma$ orbitals with moderate antibonding character to Mg. When CS is adsorbed on Fe, the bond length becomes 1.509 \AA, due to the $\pi$-electron backdonation effect of Fe on the antibonding orbitals of CS. The same occurs with CO, in which the CO bond lengths in the Mg- and Fe-adsorbed complexes are 1.117 \AA~and 1.120 \AA, respectively, shorter than that in the gas-phase molecule (1.123 \AA).  
Remarkably, the largest CS bond shortening compared to CO, together with the differences in their dipole moments (0.1 and 1.9 Debye for CO and CS, respectively), causes the BE of CS to be almost doubled with respect to CO. We previously observed the same phenomenon when characterizing the interaction of CO and CS on water ice \citep{perrero2022}, where the larger dipole of CS caused an increase in BE of 50\% with respect to the CO BE value. This further strikes the difference in the chemical behaviour of CO and CS.

Table \ref{tab:all_clusters} shows a summary of the results.
We obtained a range of values for the adsorption onto Mg centers and one value for each Fe. It appears that CS and CO prefer to interact with Fe rather than Mg, while H$_2$O clearly shows a larger affinity with Mg rather than with Fe. However, the behaviour of H$_2$S is not straightforward. The interaction with both metals seems to be comparable in clusters A and B, but cluster C represents an exception. In fact, there is a notable difference between the 74-85 kJ mol$^{-1}$ BE values computed when H$_2$S interacts with Mg and the BE value of 97 kJ mol$^{-1}$ computed for the adsorption onto Fe. An overall look at the values, particularly the BEs obtained for the Fe site at different positions (X, W, Y, and Z) hints at a possible role of the coordination of the metal center in determining the binding strength. Indeed, sites W and Z (cluster C) are the most exposed and less coordinated, and they tend to give the highest BE, while site X (cluster A) is the most coordinated and is usually the weakest one. 
In general, cluster C, in addition to being the most stable structure and offering the largest number of binding sites, is also covering the entire set of binding energies obtained with the entire set of nanoclusters. All of these reasons are in favour of choosing this structure (cluster C) for the investigation of the adsorption of the whole set of S-bearing species. 

\begin{table}
   \caption {BEs (in kJ mol$^{-1}$) of H$_2$S, H$_2$O, CS and CO adsorbed on forsterite and olivine clusters. BE ranges result from the presence of different adsorption sites on the same cluster.}
    \centering
\begin{tabular}{lccccc} 
\hline
Cluster     & Site	&	H$_2$S &	H$_2$O &	CS &	CO \\
     \hline
Fo	& Mg &	70-77 &	115-167 &	89-91 & 48-50 \\
A   & Fe &	68 &	94 &	121 &	66 \\
A	& Mg &	76-78 &	121-125 &	92-93 &	50-52 \\
B	& Fe &	73	 & 98 &	109 &	62 \\
B	& Mg &	72-74 &	119-122 &	91-93 &	49-51 \\
C	& Fe &	97 &	127 &	120 &	70 \\
C	& Mg &	74-85 &	118-131 &	90-91 &	46-54 \\
\hline
\label{tab:all_clusters}
\end{tabular}
\end{table}

\subsection{Adsorption of S-bearing species}

\begin{figure}
    \centering
    \includegraphics[width=\columnwidth]{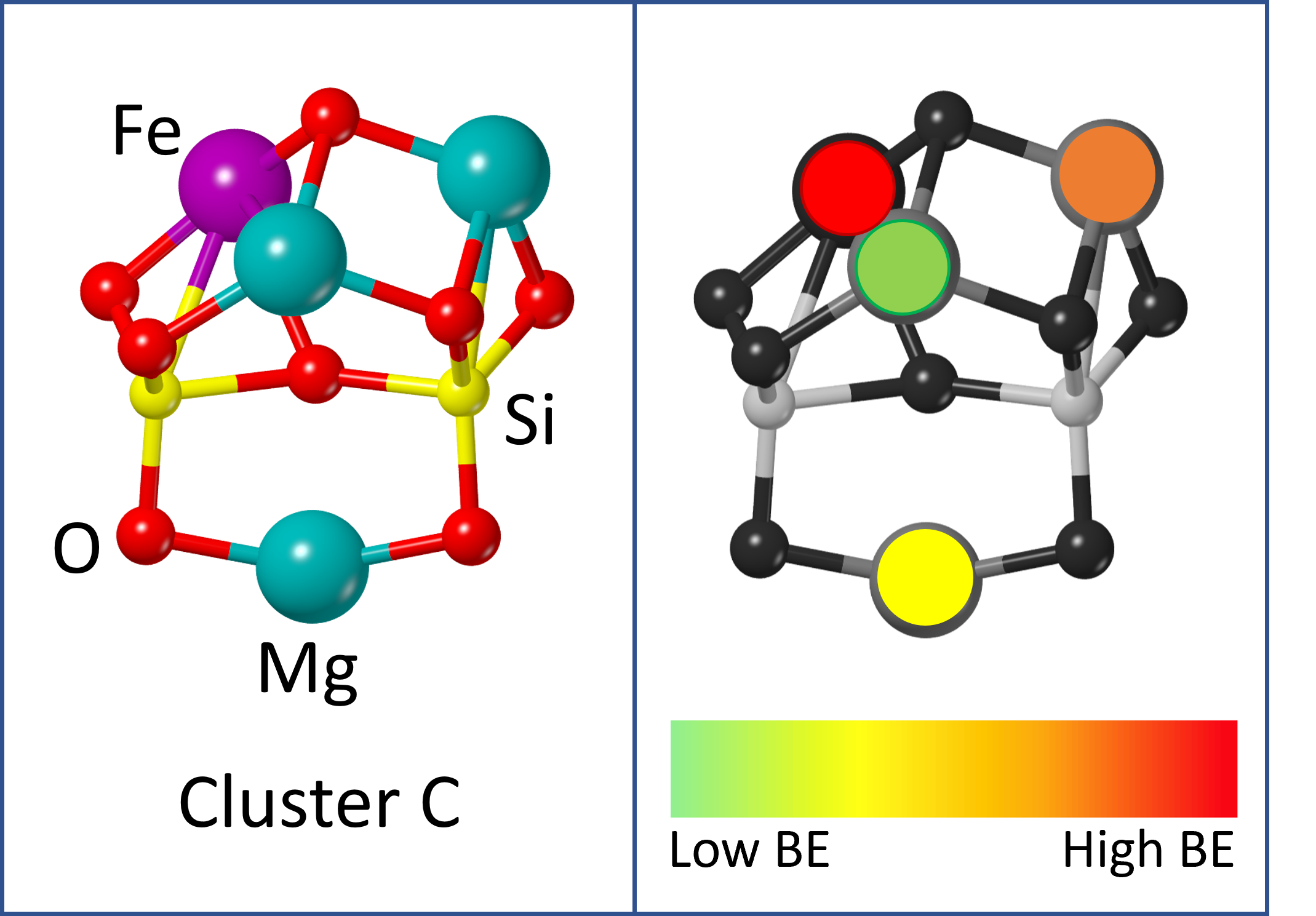}
    \caption{Colour map showing the strength of each binding site from weaker (green) to stronger (red), depending on which species has been adsorbed. It is valid for all the species characterized in this work, with the exception of \ch{SO} and \ch{SO+}, for which the Fe is the weakest binding site. In the case of H$_2$S$^+$ and HS$^+$, the cation dissociate when adsorbed onto Fe.}
    \label{fig:ads_map}
\end{figure}

The set of the adsorbed species includes CS, H$_2$S, H$_2$S$^+$, HS, HS$^+$, S, S$^+$, SO, and SO$^+$. 
These species have been selected for being the most abundant sulphur carriers in diffuse clouds and the external layers of photon-dominated regions (PDR) where the water ice has not been formed yet, either in the cloud envelopes or the dense PDRs formed in the vicinity of recently formed young massive stars \citep{fuente2003,neufeld2015, goicoechea2021, fuente2023}. We modelled four adsorption complexes for each species, from which some common characteristics emerged: i) we can draw a colour map (see Figure \ref{fig:ads_map}) to highlight the strength of the adsorption sites which is valid for all the species; ii) the \ch{Fe\bond{single}S} interaction is stronger than the \ch{Mg\bond{single}S} one (see Table \ref{tab:all_species}). Although we may attribute part of the reason to the coordination of the metal center, it is not the only explanation to such differences; iii) SO and SO$^+$ interact with the metal center through oxygen, which in this case is the most electronegative atom. This is responsible for the differences in the adsorption energies listed in Table \ref{tab:all_species}; iv) the presence of a charge dramatically increases the BE, as we observed for the four ions included in this study, (H$_2$S$^+$, HS$^+$, S$^+$, and SO$^+$).

From the calculations, we found that H$_2$S$^+$ and HS$^+$ are physisorbed on Mg, but chemisorbed on Fe, in which the proton transfers to one of the oxygen atoms belonging to the closest silicate unit (similarly to what we observed for H$_2$S on Fo), and the charge is distributed onto the entire system. Moreover, when an open-shell species adsorbs on Fe, part of its spin density transfers from the sulphur atom to the metal, increasing the binding energy as in the case of S and HS. On the other hand, this effect is prevented for SO, whose two unpaired electrons remain localized onto the radical species, due to the adsorption through the oxygen atom. The BEs of both SO and SO$^+$ can be explained by looking at the behaviour of H$_2$O, in which the interaction with the metal center takes place through oxygen and for which the interaction \ch{Mg\bond{single}O} is favoured over the \ch{Fe\bond{single}O} one. Finally, the inclusion of the $\Delta$ZPE correction in computing the BE(0) does not alter the trend derived from the analysis of the BEs.

\begin{table}
   \caption {Calculated binding energies (BE and BE(0), in kJ mol$^{-1}$) for all the considered S-bearing species on the cluster C.}
    \centering
\begin{tabular}{lcccc} 
\hline
 & \multicolumn{2}{c}{BE} & \multicolumn{2}{c}{BE(0)}  \\
	& Mg &	Fe & Mg & Fe \\
 \hline
CS & 	90-100 & 	120 & 85-96 & 115  \\
H$_2$S  & 	74-85	& 97 & 67-78 & 89  \\
H$_2$S$^+$ & 	354-364 & 	538 & 345-355 & 520 \\
HS & 	70-80	& 203 & 65-76 & 194 \\
HS$^+$ & 	498-508 & 	698 & 491-503 & 680  \\
S & 	64-87	& 175 & 63-85 & 171 \\
S$^+$	 & 470-481 & 	534 & 471-483 & 534  \\
SO & 	80-96	& 69  & 77-92 & 66 \\
SO$^+$	 & 586-616 & 499	& 579-607 &  491 \\
\hline
\label{tab:all_species}
\end{tabular}
\end{table}

\section{Discussion and astrophysical implications}\label{sec:discussione}

\begin{figure}
    \centering
    \includegraphics[width=\columnwidth]{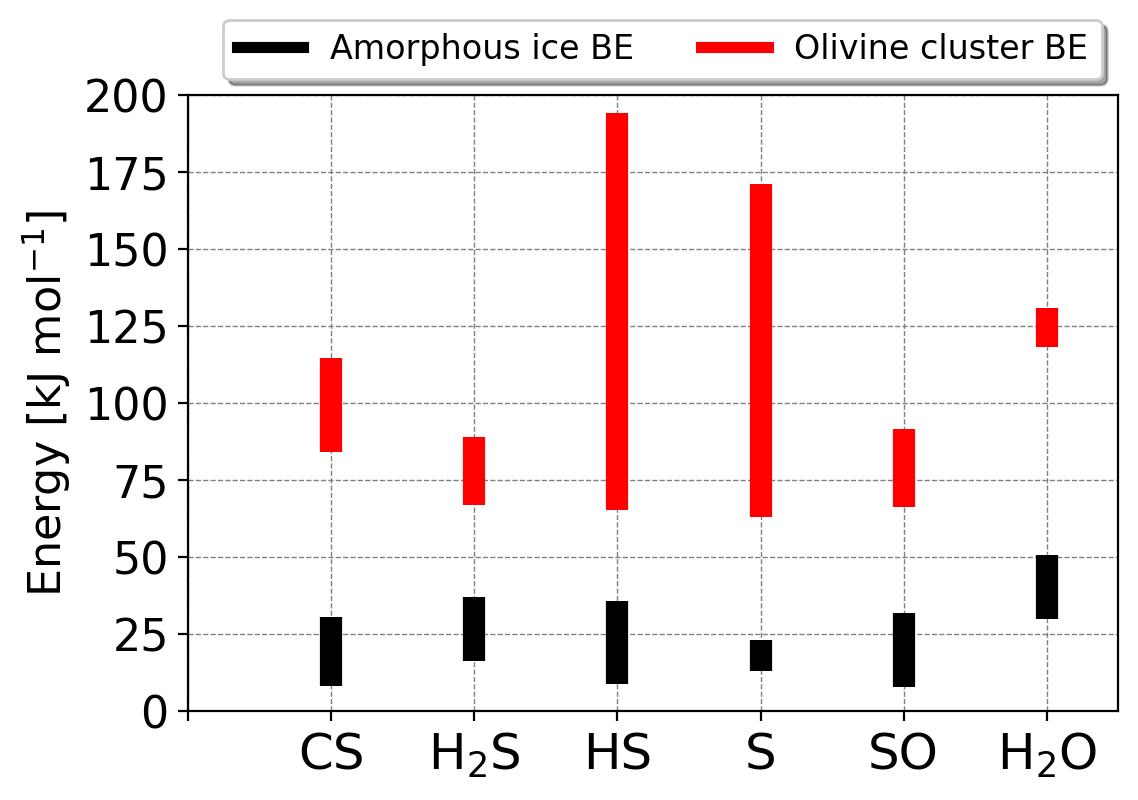}
    \caption{Binding energies (BE(0), in kJ mol$^{-1}$) ranges covered by CS,\ch{H2S}, HS, S, SO and \ch{H2O} adsorbed on amorphous ice \citep {ferrero2020,perrero2022} and on olivine.}
    \label{fig:be_bars}
\end{figure}

In a previous work by some of us \citep{perrero2022}, the BEs of 17 S-bearing species adsorbed onto crystalline and amorphous water ice were computationally studied adopting a periodic approach and the DFT//HF-3c level of theory. The B3LYP-D3 functional was used to compute the energy of closed-shell systems (such as CS and H$_2$S), while for open-shell species (such as HS, S, and SO) M06-2X was adopted. These functionals were combined with an Ahlrichs triple zeta valence quality basis set, supplemented with a double set of polarization functions \citep{schafer1992}. Although the computational approach adopted in \cite{perrero2022} differs from this work, it does not justify the large differences between the BE(0) values of the same set of species on the amorphous ice surface and the olivine nanoclusters (see Table \ref{tab:ice-vs_olivine} and Figure \ref{fig:be_bars}). While in the first case the interactions are weak and controlled mainly by the dispersion forces, BE(0)s computed in this work are large and determined mostly by electrostatic interactions, to which occasionally electron transfer phenomena add up. 

The consequences of such a strong interaction between sulphur and the mineral surface is that S-containing species have a high chance to freeze out onto the core of dust grains and not be able to desorb into the gas phase even at the temperatures found in a diffuse cloud (around 100 K). In those environments, the presence of charged gas-phase species should also favour the adsorption process as long as the grains are negatively charged \citep{cazaux2022, fuente2023}. Even if the grain particles are mostly neutrals or positively charged, the accretion of neutral sulphur species on the grain surfaces could enhance the number of sulphur atoms in refractories. On the other hand, it should be considered that dust grains in dense clouds are not completely covered by the ice mantle, therefore still exposing part of their mineral surface into the gas \citep{potapov2021}. As it clearly appears in Figure \ref{fig:be_bars}, \ch{H2O} adsorption on olivine is more energetic than \ch{H2S}, meaning that water is more likely (and probable) than hydrogen sulphide to form a stable interaction with the mineral. However, the modest interaction between sulphur-containing species and water ice allows these molecules to diffuse on its surface and likely move towards the portion of exposed core of the grains. Therefore, whether considering the physical conditions of a diffuse or those of a dense cloud, S-bearing species would have a certain chance either to directly stick onto the grain core or diffuse from the mantle towards the exposed fraction of the core grain where they become strongly chemisorbed.

\begin{table}
   \caption {Binding energies (BE(0), in kJ mol$^{-1}$) of five S-bearing species and water on amorphous ice \citep {ferrero2020,perrero2022} and on olivine.}
    \centering
\begin{tabular}{lcc} 
\hline
Species & BE(0) on amorphous ice &	BE(0) on olivine   \\
 \hline
CS & 8.3-30.6 & 84.4-114.6 \\
H$_2$S & 16.4-37.3 & 67.1-89.0  \\
HS & 9.0-35.9 & 65.4-194.4 \\
S & 13.1-23.3 & 63.0-171.2 \\
SO & 7.8-32.0 & 66.3-91.8 \\
\hline
H$_2$O & 30.0-50.8  & 118.2-131.0   \\
\hline
\label{tab:ice-vs_olivine}
\end{tabular}
\end{table}

Another interesting issue is the tendency of certain species, namely H$_2$S, HS$^+$, and H$_2$S$^+$, to chemisorb onto olivine nanoclusters, planting the seed for the formation of the \ch{Fe\bond{single}S} bond present in sulphide minerals such as troilite and pentlandite. In the work of \cite{kama2019}, approximately 90\% of the sulphur is predicted to be locked in the refractory residue of protoplanetary disks, the main carriers being precisely sulphide minerals, followed by sulphur chains S$_n$, through to the study of the photospheres of a sample of young stars. However, chemical models \citep{druard2012} and laboratory experiments \citep{woods2015} predict a scarce abundance of sulphur chains in these environments, reason why sulphide minerals are considered the major reservoir of refractory sulphur in the protoplanetary disk. These two minerals are also found in cometary dust and meteoritic rocks \citep{kallemeyn1989,jm2012meteoritos}. From a thermodynamic point of view, \ch{Fe\bond{single}S} is also more favoured than \ch{Mg\bond{single}S}, as the latter is only found in the less common enstatite chondrites, that due to their formation in reducing conditions, are rich in alkaline sulphides like the Mg-containing niningerite and alabandite \citep{sears1982}.

In this work we did not consider charged dust grains, even though the model of \cite{ibanez2019} predicted them to be positively charged in a diffuse medium and negatively charged in denser environments. If this was the case, the interaction between gas-phase cations and positively charged grains in diffuse clouds would clearly be hampered by electrostatic repulsion. However, our results concerning positively charged sulphur-bearing species adsorbed on neutral dust grain cores hold true for the case in which neutral gas-phase species interact with charged dust grains. Although the atomistic details of the system should vary due to the presence of a charge on the olivine nanocluster, the interaction energies should be of the same order of magnitude, since the presence of a charge in the system would still be the main driving force of the interaction. On the other hand, \cite{fuente2023} argue that sulphur would remain ionized in the gaseous medium until grains are expected to be negatively charged. If this were the case, the electrostatic interactions would be responsible for an increased sulphur depletion effect.

According to what \cite{hilyblant2022} suggested, the fact that sulphur depletion increases progressively as the core evolves could be in agreement with the fact that during the evolution, atomic S freezes out on the surface of grains and, depending on which surface it reaches, it can either become part of the mineral core or as \cite{cazaux2022} propose, it can fall on the ice mantle and be subjected to photoprocessing, thus reacting and forming a refractory residue.

Our calculations are also relevant to undestand the sulphur chemistry in the dense PDRs formed on the molecular cloud/HII regions interfaces. It has been considered that in these dense warm regions (dust temperature $\sim$ 100 K), the chemistry is well explained with gas-phase networks since freeze-out on grain surfaces is negligible. Our new calculations suggest that this may not be the case since the binding energies of sulphur species is up to a factor of $\sim$10 in bare grains. This means that grain-gas interactions need to be considered in these warm environments.

Finally, the assumption that S-bearing species become part of the grain core, rather than interacting with the icy mantles, would be in agreement with the fact that they are usually used as shock tracers \citep[e.g.,][]{sato2022,zhang2023}. In shock regions, which are caused by the impact of stellar outflows with the quiescent surrounding gas, H$_2$S can be sputtered off the grains and enter in the gas phase, where it marks the zone that has been affected by such an impact \citep{woods2015}. The harsh conditions of the shock would therefore provide enough energy to break the strong interaction between sulphur and the mineral surface, allowing its transformation in H$_2$S and successive ejection in the gas phase \citep{holdship2017}.

\section{Conclusions}\label{sec:conclusioni}
In this work, we studied the interaction of nine S-bearing species, CS, H$_2$S, H$_2$S$^+$, HS, HS$^+$, S, S$^+$, SO, and SO$^+$, which include both neutral and charged, as well as closed-shell and open-shell systems. We characterized their adsorption complexes and computed the corresponding binding energies on  olivine nanoclusters (forsterite and the Fe-containing analogues) by means of quantum mechanical calculations, to simulate the interaction of a gas-phase species with the surface of a silicate bare dust grain in the diffuse interstellar medium. Our main findings are that: i) from a thermodynamic point of view, the \ch{Fe\bond{single}S} interaction is more stable than \ch{Mg\bond{single}S}, while when the oxygen end is interacting with the metal center we notice the opposite trend; ii) S-bearing species are adsorbed much more strongly on the core of bare dust grains than on their icy mantles (represented by pure amorphous water ice surfaces; iii) the metal centers of the olivine clusters can be responsible for the dissociation of hydrogenated S-bearing species, especially if the latter are positively charged; iv) the dissociation of \ch{H2S} on forsterite cluster presents a small barrier, that is likely to be overcome when considering the energy liberated by the adsorption process. 
The implications of such findings are that sulphur species have a high probability of sticking onto the surfaces of bare dust grains and becoming part of their core in the form of refractory materials, as predicted by a number of experimental and observational studies, in addition to chemical models. Indeed, more investigations are needed in order to finally unveil sulphur chemistry in the ISM. However, the trapping of sulphur into the grain cores in the early stages of the evolution of a prestellar core could, in part, account for the S-depletion in dense cores.  

\section*{Acknowledgements}

This project has received funding within the European Union’s Horizon 2020 research and innovation programme from the European Research Council (ERC) for the projects ``Quantum Chemistry on Interstellar Grains” (QUANTUMGRAIN), grant agreement No. 865657 and ``The trail of sulphur: from molecular clouds to life” (SUL4LIFE), grant agreement No. 101096293, and from the Marie Sklodowska-Curie for the project ``Astro-Chemical Origins” (ACO), grant agreement No 811312. The Italian Space Agency for co-funding the Life in Space Project (ASI N. 2019-3-U.O), the Italian MUR (PRIN 2020, Astrochemistry beyond the second period elements, Prot. 2020AFB3FX) are also acknowledged for financial support. Authors (J.P.) acknowledge support from the Project CH4.0 under the MUR program "Dipartimenti di Eccellenza 2023-2027" (CUP: D13C22003520001).
The Spanish MICINN is also acknowledged for funding the projects PID2021-126427NB-I00 (A.R.), PID2019-106235GB-I00 (A.F.), and PID2020-116726RB-I00 (L.B.-A.). We also thankfully acknowledge the computer resources and assistance provided by the Barcelona Supercomputing Center (BSC) and CSUC.

%%%%%%%%%%%%%%%%%%%%%%%%%%%%%%%%%%%%%%%%%%%%%%%%%%
\section*{Data Availability}

The data underlying this article are freely available in Zenodo at \url{https://zenodo.org/10.5281/zenodo.8363882}.



\bibliographystyle{mnras}
\bibliography{my_bib} % if your bibtex file is called example.bib

%%%%%%%%%%%%%%%%%%%%%%%%%%%%%%%%%%%%%%%%%%%%%%%%%%

%%%%%%%%%%%%%%%%% APPENDICES %%%%%%%%%%%%%%%%%%%%%

% Don't change these lines
\bsp	% typesetting comment
\label{lastpage}
\end{document}